\documentclass{ws-mpla}
\usepackage[super]{cite}
\usepackage{hyperref}
\usepackage[pdftex]{graphicx,xcolor}

\newcommand{\bfi}[1]{\mbox{\boldmath $#1$}}
\newcommand{\vrr}{{\bfi r}}
\newcommand{\vK}{{\bfi K}}
\newcommand{\vR}{{\bfi R}}
\newcommand{\vs}{{\bfi s}}

\begin{document}

\markboth{Matsuzaki and Yahiro}{Neutron skin  of $^{48}$Ca}

%%%%%%%%%%%%%%%%%%%%% Publisher's Area please ignore %%%%%%%%%%%%%%
\catchline{}{}{}{}{}
%%%%%%%%%%%%%%%%%%%%%%%%%%%%%%%%%%%%%%%%%%%%%%%%%%%%%%%%%%%%%%%%%%%

\title{Neutron skin  of $^{48}$Ca deduced from interaction cross section}

\author{Masayuki Matsuzaki}

\address{Department of Physics, Fukuoka University of Education,
Munakata, Fukuoka 811-4192, Japan\\
matsuza@fukuoka-edu.ac.jp}

\author{Masanobu Yahiro}

\address{Department of Physics, Kyushu University, Fukuoka 819-0395, Japan\\
orion093g@gmail.com}

\maketitle

\pub{Received (Day Month Year)}{Revised (Day Month Year)}

\begin{abstract}
The neutron skin thickness of $^{48}$Ca was deduced from the interaction cross section 
by adopting a microscopic optical potential. The optical potential used was constructed by 
folding a chiral $g$ matrix and the Skyrme mean-field densities renormalized by 
considering the information of the interaction cross section. 
The result was $R_{\rm skin}$ = 0.139$\pm$0.058 fm.

\end{abstract}

\ccode{PACS Nos.: }

The neutron skin thickness $R_{\rm skin}$ is not only one of the basic quantities of the structure 
of terrestrial nuclei, but also strongly correlated with the stiffness of the equation 
of state of the nucleonic matter that composes neutron stars. This can be deduced in 
several ways. Among them, parity-violating electron scattering~\cite{PRC.63.025501} is thought to be 
the most precise means of determining the neutron root-mean-square (RMS) radii $R_{\rm n}$, which 
are hardly determined by hadronic probes. In contrast, the proton RMS radii $R_{\rm p}$ were accurately determined 
from the elastic electron scatterings. 
Theoretically, only mean-field calculations are practically available for heavy 
nuclides. Hamiltonians or energy-density functionals adopted there contain many 
parameters informed by the measured quantities of representative stable and some unstable 
nuclides. This indicates that the neutron sector is less constrained, 
particularly for heavy nuclides. Mean-field calculations directly give $R_{\rm n}$, $R_{\rm p}$, and 
consequently $R_{\rm skin}=R_{\rm n}-R_{\rm p}$, but the results should be critically assessed. 

A new method that relies more directly on another experimental 
observable, the reaction cross section $\sigma_{\rm R}$, was proposed in Refs.~\cite{Tagami:2020bee,PhysRevC.104.054613}. 
From the experimental perspective, this is a method to extract $R_{\rm skin}$ from $\sigma_{\rm R}$ based on 
a reaction model with a microscopic optical potential. At the same time, from a theoretical perspective, 
this improves the calculated $R_{\rm skin}$ given by mean-field models with energy-density functionals 
whose parameters might not yet be fully constrained. 
Specifically, the adopted optical potential is constructed 
by folding a chiral $g$ matrix~\cite{Toyokawa:2017pdd}, given by localizing the one originally constructed by taking 
into account the next-to-next-to-next-to 
leading order (N$^{3}$LO) two-body force and the NNLO three-body force in chiral 
perturbation~\cite{PRC.88.064005}, 
and Gogny/Skyrme mean-field densities. 
As a result, the authors 
of Refs.~\cite{Tagami:2020bee,PhysRevC.104.054613} 
obtained for $^{208}$Pb, 
$R_{\rm skin}=$ 0.278 $\pm$ 0.035 fm$\,$~\cite{Tagami:2020bee} and 
$R_{\rm skin}=$ 0.416 $\pm$ 0.146 fm$\,$~\cite{PhysRevC.104.054613}, respectively,  
which are 
consistent with PREX II with parity-violating electron scattering, $R_{\rm skin}=$ 0.283 $\pm$ 0.071 fm$\,$~\cite{Adhikari:2021phr}. 

In the present study, we examine $^{48}$Ca by adopting a Skyrme parameter set. 
Reference data were obtained from Tanaka et al.~\cite{Tanaka:2019pdo}. They measured the interaction 
cross sections $\sigma_{\rm I}$ of Ca isotopes + $^{12}$C scatterings at 280 MeV/nucleon, 
which are almost the same as $\sigma_{\rm R}$ above 100 MeV/nucleon. 
We compare in the following 
our results for $R_{\rm skin}$ with theirs, $R_{\rm skin}$ = 0.146 $\pm$ 0.060 fm, deduced 
using the optical limit of the Glauber model with the Woods-Saxon density. 
A dip in the isotope dependence was observed for $^{48}$Ca, whereas the theoretical 
result was smooth~\cite{Tagami:2019svt}. In addition, a precision datum from the ongoing CREX project 
will soon be obtained. 

\begin{figure}[ph]
\centerline{\includegraphics[width=3.0in]{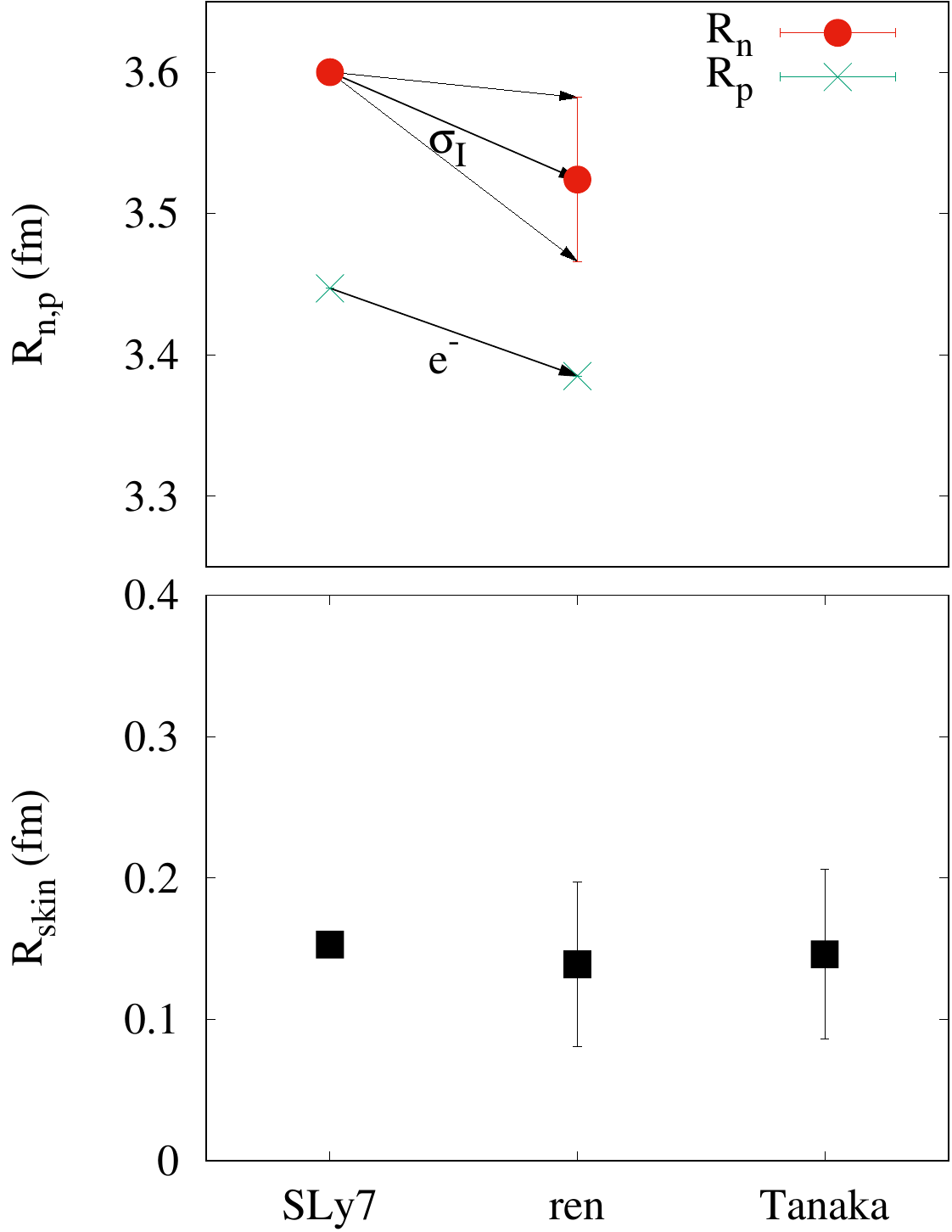}}
%\vspace*{10pt}
\caption{Neutron and proton radii, and skin thicknesses: Directly given by a Skyrme-Hartree-Fock-Bogoliubov 
calculation ("SLy7": left), deduced in the present renormalization method ("ren": center), and given 
by Tanaka et al.~\cite{Tanaka:2019pdo} adopting the Glauber model ("Tanaka": right). 
Effects of the density renormalization to reproduce the electron (e$^{-}$) scattering and the interaction 
cross sections ($\sigma_{\rm I}$) data are shown by arrows.\protect\label{fig1}}
\end{figure}

We adopted the SLy7 parameter set, which was constructed by improving the famous SLy4 set~\cite{Chabanat:1997un}. 
A Skyrme-Hartree-Fock-Bogoliubov (SHFB) calculation~\cite{Schunck:2016uvm} using this parameter set directly 
yields ($R_{\rm n}$, $R_{\rm p}$, $R_{\rm skin}$)=(3.600, 3.447, 0.153) fm. 
In contrast to the $^{208}$Pb case~\cite{PhysRevC.104.054613}, the third one , $R_{\rm skin}=$ 0.153 fm, 
was consistent with the reference data of Tanaka et al., $R_{\rm skin}$ = 0.146 $\pm$ 0.060 fm. 
In order to look into the results more closely, 
we consulted the precision electric scattering data adopted there, $R_{\rm p}$ = 3.385 fm, 
and renormalized the SHFB densities to remedy possibly weak constraints on mean-field parameters. 
The theoretical framework is briefly summarized here. 
The optical potential to determine the scattering wave function and thus the cross section 
is given by folding the mean-field densities and the $g$ matrix and consists 
of the direct and exchange parts, 
\begin{eqnarray}
U^{\rm DR}(\vR) \hspace*{-0.15cm} &=& \hspace*{-0.15cm} 
\sum_{\mu,\nu}\int \rho^{\mu}_{\rm P}(\vrr_{\rm P}) 
            \rho^{\nu}_{\rm T}(\vrr_{\rm T})
            g^{\rm DR}_{\mu\nu}(s;\rho_{\mu\nu}) d \vrr_{\rm P} d \vrr_{\rm T}, \\
U^{\rm EX}(\vR) \hspace*{-0.15cm} &=& \hspace*{-0.15cm}\sum_{\mu,\nu} 
\int \rho^{\mu}_{\rm P}(\vrr_{\rm P},\vrr_{\rm P}-\vs)
\rho^{\nu}_{\rm T}(\vrr_{\rm T},\vrr_{\rm T}+\vs) \nonumber \\
            &&~~\hspace*{-0.5cm}\times g^{\rm EX}_{\mu\nu}(s;\rho_{\mu\nu}) \exp{[-i\vK(\vR) \cdot \vs/M]}
            d \vrr_{\rm P} d \vrr_{\rm T},~~~~
\end{eqnarray}
where $\vs=\vrr_{\rm P}-\vrr_{\rm T}+\vR$ 
for the coordinate $\vR$ between the projectile (P) and target (T). The coordinate 
$\vrr_{\rm P}$ 
($\vrr_{\rm T}$) denotes the location of the interacting nucleon 
measured from the center of mass of P (T). 
Each of $\mu$ and $\nu$ stands for the $z$-component of isospin. 
Note that we use the localized version of $U^{\rm EX}$. 
The $g$ matrix depends on the local density at the midpoint of the interacting nucleon pair, 
and taken from the numerical table~\cite{localized}. 
The densities in the above potentials are renormalized as: 
We define the scaled density $\rho_{\rm scaling}(\vrr)$ from the original density $\rho(\vrr)$ as
\begin{equation}
\rho_{\rm scaling}(\vrr)=\frac{1}{\alpha^3}\rho(\vrr/\alpha)
\end{equation}
with a scaling factor
\begin{equation}
\alpha=\sqrt{ \frac{\langle \vrr^2 \rangle_{\rm scaling}}{\langle \vrr^2 \rangle}} .
\end{equation}
The actual procedures to determine $\alpha$ (of p and n) are as follows: 
First, we scale the proton density so as to be 
$R_{\rm p}({\rm scaling})=R_{\rm p}({\rm exp})$ ; 
second, we scale the neutron density so that the calculated $\sigma_{\rm I}$ reproduces the data with an error bar, 
as shown in Fig~\ref{fig1}. 

%%%%%%%%%%%%%%%%%%%%%%%
%%%  Figure
%%%%%%%%%%%%%%%%%%%%%%%

Double-folding calculations with renormalized densities yield 
($R_{\rm n}$, $R_{\rm skin}$)=(3.524$\pm$0.058, 0.139$\pm$0.058) fm. 
This indicates that the decreases in $R_{\rm n}$ and $R_{\rm p}$ cancel each other, and consequently 
$R_{\rm skin}$ remains similar as shown in Fig~\ref{fig1}. 
Therefore the renormalization is necessary but the difference in the adopted reaction models does not appear here. 
In order to see the influence of the adopted mean field, first we checked the standard 
SLy4 set. The reason why we adopted the SLy7 set is that it is advertised that the SLy7 set was obtained by 
improving the SLy4 with respect to both a spin-gradient term and a more refined two-body 
center of mass correction and the joint contribution of the two terms 
brings significant improvement for Pb isotopes\cite{Chabanat:1997un}. But we found that these effects on the present calculation were 
negligible. Second, we examined the Gogny D1S force instead of Skyrme forces. 
It lead to $R_{\rm skin}=$ 0.105 $\pm$ 0.06 fm. 
As for the effective nucleon-nucleon interaction, only the adopted chiral $g$ matrix is available for us. 
We think it reliable because it was confirmed not only in scattering calculations~\cite{Toyokawa:2017pdd} 
but also in structure calculations~\cite{PhysRevC.100.034310}.

Finally we compare the present result with other information: Results of 
the high-resolution E1 polarizability experiment, 0.17$\pm$0.03 fm~\cite{Birkhan:2016qkr}, and an 
{\it ab initio} coupled-cluster calculation available for light nuclides, 
0.135$\pm$0.015 fm~\cite{Hagen:2015yea}. We confirmed that all the results examined here fall 
into these ranges around 0.15 fm. 
By consulting the fitted correlation between $R_{\rm skin}$ of $^{48}$Ca and 
$^{208}$Pb $\,$~\cite{Tagami:2020shn} 
\begin{equation}
R_{\rm skin}^{48}=0.5547\,R_{\rm skin}^{208}+0.0718
\end{equation}
and that between $R_{\rm skin}^{208}$ and the slope parameter $L$ of symmetry energy~\cite{Roca-Maza:2011qcr} 
\begin{equation}
R_{\rm skin}^{208}=0.00147\,L+0.101\, ,
\end{equation}
this suggests rather soft slope parameters around 27 MeV, 
in contrast to the $^{208}$Pb results of PREX II and Refs.~\cite{Tagami:2020bee,PhysRevC.104.054613}.

\bibliographystyle{ws-mpla}

\end{document}